# A class where qualitative discussions, coming weeks before computationally complicated practice, helps students' problem solving abilities

D. J. Webb, Department of Physics, University of California, Davis, CA

**Introduction**

Psychologists have long known that an expert in a field not only knows significantly more individual facts/skills than a novice but also has these facts/skills are organized into a mental hierarchy that links the individual facts (at the bottom of the hierarchy) together with larger more-encompassing ideas (the top of the hierarchy).  This expert cognitive structure allows an expert physicist to "see" a physical situation differently than a novice (understand the important features of the physical situation), analyze the physical situation differently (more effectively use qualitative thinking), and so solve problems more efficiently.  Educators might view their problem as how to structure a curriculum to most efficiently help students develop this expertise.  In the Spring quarter of 2012, UC Davis offered 4 sections (about 180 students each) of the first quarter of introductory physics, Physics 9A, covering Newtonian mechanics.  In this paper, I discuss some curricular details of one section (the treatment group) that had the entire 10-week quarter's set of ideas introduced, largely qualitatively, in the first 6 weeks followed by 4 weeks where students learn to use those ideas to solve the algebraically complicated problems that physicists prize.  The other three sections of 9A were organized in the usual way by introducing new ideas almost simultaneously with the algebraically complicated problems, one topic at a time, throughout the 10-week quarter.  The treatment group as well as one of the other three sections were identical except for content organization so together they constitute a controlled study.  After controlling for a student's overall academic ability (using GPA) as well as initial understanding of Newtonian mechanics (using the Force Concept Inventory[1], FCI), the treatment group was found, with better than 99% confidence, to score higher on a final exam that was completely blind to the instructor.  This result is potentially important because almost any physics class could be organized along these lines of first letting students build a more expert mental framework and then asking them to use that framework and, in doing so, strengthen it.

For physics the lowest level of the mental hierarchy of an expert presumably includes all of the specific detailed physical processes and situations that the expert remembers (or can re-construct) and the highest level includes the main overarching models that physicists use to give both qualitative and quantitative explanations of physical processes.  For the intro-physics classes discussed in this paper, the overarching models include Galilean relativity, Newtonian mechanics, conservation of energy, conservation of momentum, and conservation of angular momentum.  Students need to learn the overarching models as well as how to connect them to specific phenomena that are new to the student.  In other words, the student is supposed to construct much of the major mental framework of an expert. A student initially learns the meaning of these ideas by seeing the words and equations defined and used in discussions of various physical phenomena. Then they strengthen their understanding through practice in qualitative and quantitative analysis of either simple or complex physical situations. Examination of the "difficult" problems in almost any standard introductory-physics textbook suggests that physicists especially prize a student's ability to analyze a complicated physical situation using one or more of these overarching physical models to derive a quantitative result. For the purposes of this paper, I accept that success at solving these standard "hard" problems is one of the distinguishing abilities of an expert in the field.

The last 30 years research in the field of Physics Education has shown[2] that standard lectures, problem solving, and performing standard laboratory experiments do not lead to excellent student understanding of highest level of ideas in the hierarchical mental structure of an expert physicist. This research has also shown[3] how to dramatically increase our students' understanding of the highest hierarchical level, largely by devoting class-time specifically to student-student discussions of these big ideas and their qualitative application so that students can work very directly toward building an expert conception of the ideas. Unfortunately, to date the research also generally shows that a significantly increased understanding of the ideas that make up the highest level of the expert's mental hierarchy does not usually result in significantly better performance on the standard problems that physicists prize. If an expert's excellent performance is due to their understanding and use of these big ideas, then it is something of a puzzle that a student who understands the big ideas does not seem to succeed in using this understanding (in a way that we can measure) to produce more expert performance. The treatment class discussed in this paper was designed to help us understand this puzzle.

**Experiment**

The author was the instructor of two of the classes that were offered in Spring 2012, the treatment class (section I) and one of the control classes (section II). These two sections not only had the same instructor and textbook (*University Physics* by Young and Freedman, 13th edition) but the same homework problems, lecture questions, discussion section questions, and laboratory experiments so the curriculum was essentially identical except for the organization of material. In fact, the other two lecture sections (III and IV) also used the same textbook and had the same laboratory experiments. In this paper I will try to give a complete enough picture of the curriculum of the treatment class so that other instructors can try the same experiment if they wish. Then I'll describe some of the data showing that the treatment group outperformed the control group.

In a physics class, instruction often proceeds in the following generic way. Most physics instructors probably i) begin a topic by either using an experimental result to motivate the definitions of some new variables and equations or by using results from previous learning to derive a new equation that motivates the definition of new variables, ii) work with their students to help the students understand the new ideas and their (sometimes complicated) relations to the physical world, and iii) work with their students to help them put these ideas to use in calculating interesting physical quantities or in explaining known results or in analyzing new physical situations or experiments. In a standard physics class, these three steps often proceed one-right-after-the-other topic after topic for the entire time of the class (maybe with a little time for recap and reflection at the end). The treatment group discussed in this paper basically takes steps i) and ii) for each topic and puts them into the first 60% of the school quarter and then puts step iii) in the final 40% of the quarter. For example, the syllabus for the treatment group is shown in Table 1.

A new topic in either section I or section II began with a discussion of ideas by the author (motivated either by a demonstration or by a discussion of a real world situation). The author uses multiple-choice and True-False questions during lecture and the students submit their answers using a personal-response-device (a "clicker"). After each 10-20 minutes of lecturing would be time for 2-5 clicker questions that test the students understanding of the basic concepts and/or help the students to build this basic understanding. Sometimes the students are polled before they can discuss the issue between themselves and then polled again after they have discussed it (a la Mazur's prescription[4] from Peer Instruction) and sometimes they talk to each other about the problem before they are polled. As examples of some questions asked during lecture, Figure 1 shows two of these. Figure 2 shows an example of a qualitative homework problem that still requires considerable thought

| Week | Topics | Chapters |
|---|---|---|
| 1 | **Motion** (position, velocity, acceleration) | 1,2,3 ideas |
| 2 | **Newton's Laws** (forces can change motion) | 4,5 ideas |
| 3 | **Energy Conservation** | 6,7 ideas |
| 4 | **Momentum Conservation, Rotational motion** | 8,9 ideas |
| 5 | **Rotational motion** (energy, torque, angular momentum) | 10,11 ideas |
| 6 | **Newton's Law of Gravity, Oscillations** | 13,14 ideas |
| 7 | All of the above | 1-11,13,14 Review and Calculations |
| 8 | All of the above | 1-11,13,14 Review and Calculations |
| 9 | All of the above | 1-11,13,14 Review and Calculations |
| 10 | All of the above | 1-11,13,14 Review and Calculations |

Table 1 – Syllabus for treatment section.

The picture to the right shows a hand pushing on block B which results in the spring being compressed a distance x to the left. Blocks A and B are resting on a table. 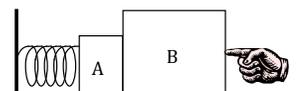

3. **The spring:**
A) interacts with block B with a significant contact force.
B) interacts with block B with a significant long-range force.
C) has no significant interaction with block B.

6. **A) for True or B) for False.** A box rests on the floor. The upward force exerted by the floor surface on the box forms a Newton's 3rd Law pair with the force exerted downward on the box by the Earth's gravity.

FIG. 1. Two of the 8 questions that were asked during the first 80 min. lecture on Newton's Laws.

and was asked during the qualitative 6 weeks of the schedule shown in the syllabus.

In lecture section II after the ideas are discussed it is time to put them into action (step iii) above) but in section I this step will wait for a few weeks. When in calculate mode I again try to keep the class interactive by asking them to work together to first think qualitatively about one or more homework problems. Figure 3 shows an example of some qualitative questions designed to help the students get started (in student-student discussion during lecture) on one of the homework problems that involve more complicated computations. I often gives my students quotations from famous expert physicists to help motivate the reasons for my teaching methods and so at this point I might refer to WHEELER'S FIRST MORAL PRINCIPLE[5] – *"Never make a calculation until you know the answer."*

**2) Hold the blocks up**

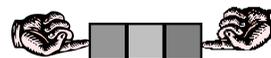

The picture to the right shows two fingers holding up three identical mass blocks by pushing them together. The fingers are exerting exactly opposing normal forces.

**a)** Sketch a reasonable free-body diagram for the **center block**. Do you need to know where the forces are placed or not (in other words, will torques matter here)?

**b)** Sketch reasonable free-body diagrams for the other two blocks. Do you need to know where the forces are placed or not (in other words, will torques matter here)? Go back and fix your diagram from part **a)** if you need to.

**c)** Will the normal force applied by the fingers need to change as you add more and more blocks? Explain.

FIG. 2. A qualitative homework question asked after initial discussions of torque.

**Think About Homework #14 Problems**

**1)** One end of a uniform meter stick is placed against a vertical wall. The other end is held by a lightweight cord that makes an angle $\theta$ with the stick. The coefficient of static friction between the end of the meter stick and the wall is 0.35. You want to understand how the angle $\theta$ affects forces required to keep the meter stick in equilibrium (for instance, how large can $\theta$ be?). Which of our tools/rules should you reach for? Sketch a free-body diagram (just draw a dot to represent the meter stick) and also an "extended" free-body diagram (use a stick to represent the stick). Can $\theta$ be (almost) as large as $\pi/2$? If not, why not?

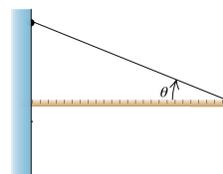

Later questions in this problem are concerned with forces required to stabilize the meter stick after hanging a mass from it. Sketch two (extended) free-body diagrams for this meter stick with the mass: one with the mass and meter stick as two separate objects and another with the mass and meter stick considered as a single object.

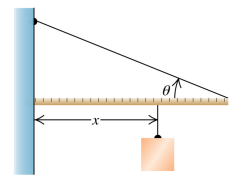

FIG. 3. A sample set of qualitative thinking that could be done before doing the specific homework problem.

## Results

The final exam was written by the two instructors from sections III and IV and another instructor. Each of these instructors has taught the course several times in the past few years. The author had no input on the final exam and did not see it until all instruction (including review sessions) had ended. Each of the four lecture sections took the same final exam at the same time. Each of the eight problems on the final exam was worth 20 points. The exam is shown in Figure 4. Each problem of the final exam was graded by two graders, one who graded sections I and II (mixed together) and one who graded sections III and IV (mixed together). Each pair of graders worked together (almost all of them sat side-by-side during their grading) to normalize their grading scale. They discussed different types of student errors and worked to award the same number of points to the same types of answer. The instructor from section III supervised the details of the grading including what was considered to be a correct solution.

**[1] Circle the answer that correctly completes the sentence correctly.**
(a) A truck is accelerated in the forward direction by the force of the ( truck's tire on the ground | ground on the truck's tire | none of the above ).
(b) A rough surface can exert a(n) ( normal | tangential | all of the above | none of the above) force.
(c) Many of the great rivers in the world have a tendency to flow from the pole towards the equator. Over time as they carry sediment towards the equator, the length of the earth's day will ( increase | decrease | not change ).
(d) For the graph of the potential energy of a particle as a function of x, the particle will have the greatest speed where the potential energy graph is ( maximum | | steepest | 0 | minimum | none of the above )
(e) For an object in circular motion ( centripetal | tangential | all of the above | none of the above ) acceleration will change the direction of the velocity vector.

Statements **(f)** through **(h)** refer to the figure below, which shows the position of an object as a function of time. Circle all the times that complete the statement correctly.

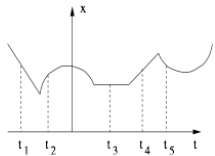

(f) The total force acting on the object is positive at ( $t_1$ | $t_2$ | $t_3$ | $t_4$ | $t_5$ ).
(g) The total force acting on the object is negative at ( $t_1$ | $t_2$ | $t_3$ | $t_4$ | $t_5$ ).
(h) The total force acting on the object is zero ( $t_1$ | $t_2$ | $t_3$ | $t_4$ | $t_5$ ).

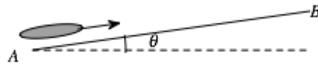

**[2]** A 90,000 kg jet has two engines which produce a constant thrust of 110,000 N each during the takeoff roll along the runway. The takeoff speed is 210 km/hr. Neglect air and rolling resistance, and assume the engine thrust is parallel to the runway. The runway is slightly inclined at $\theta = 1°$.
(a) Determine the length s of the runway required first for an uphill takeoff direction from A to B,
(b) and second determine the length s for a downhill takeoff direction from B to A.

**[3]** The figure to the right shows the potential energy $U(x)$ as a function of the position x.
(a) Mark the positions where the force is zero.
(b) What is the direction of the force at x = 4 m?
(c) If an object moving in this potential energy has a total energy of 5 Joules, mark the turning points of its motion.
(d) Assuming the total energy is 5 J, estimate the object's kinetic energy at points x = - 3 m and x = 2 m.

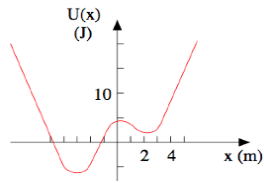

**[4]** A 50 kg woman stands up in an 80 kg canoe 5.00 m long. She walks from a point A which is 1.00 m from one end to a point B which is 1.00 m from the other end. If you ignore resistance to motion of the canoe in the water, how far does the canoe move during this process?

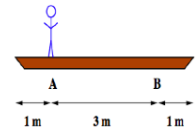

**[5]** A children's merry-go-round consists of a thin disk of mass $2m$ and radius $r_0$. It is initially spinning with angular speed $\omega_0$. There is a thin girl of mass $m$ standing at the outer edge of the merry-go-round. She walks along a radial path to the center of the merry-go-round.
(a) Explain what happens to the girl-merry-go-round system. The moment of inertia of a disk is $I = \frac{1}{2}Mr^2$.
(b) What is the final angular speed of the merry-go-round?

**[6]** The moment of inertia of a solid disk is $I = \frac{1}{2}Mr^2$. The moment of inertia of a ring is $I = Mr^2$. A disk and a ring with equal mass (m) and equal radii, $r$, both roll up an inclined plane. They start with the same linear velocity, $v$, for the center of mass.
(a) Clearly explain which will go higher.
(b) Use conservation of energy to determine the maximum vertical height, $h_d$, the disk and the maximum vertical height, $h_r$, the ring will reach.

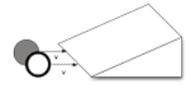

**[7]** An art object of nonuniform composition is suspended horizontally from the ceiling by two wires at its ends. The object is of length $L$, and its center of mass is $L/4$ from its left end. The wire at its right end makes a 30° angle with the horizontal and has a 100 N tension. Determine the object's weight and the tension in the other wire.

**[8]** The density of a typical asteroid is $2.5 \times 10^3$ kg/m$^3$. This is important, for you're intent on finding the largest heavenly body on which you can stand and launch a rock into a low circular orbit *by hand*, and your best throw is only 40 m/s.
(a) Assuming a spherical asteroid (not a very realistic assumption), what is the radius of the asteroid you desire? (Note: "low orbit" means the minimum possible radius. The volume of a sphere is $\frac{4}{3}\pi R^3$.)
(b) If you threw the rock instead straight up, would it go forever, and if not, how high would it go? Assume the radius is for the asteroid in part **(a)**.

FIG. 4. This final exam was given to all four lecture sections offered at UC Davis in Spring 2012.

The analysis of the data are discussed in more detail in Ref. 6. As discussed there, after controlling for general academic ability (using the student's GPA) as well as for understanding of Newtonian physics upon entering the class (using the FCI given at the beginning of the quarter), and previous experience in physics classes, the treatment group performed better on the final exam. The exam gain by the treatment group (section I) was found to be 6.2 ± 2.3 points out of the 120 total points on the final exam. That this number is 2.7 standard deviations away from a null result tells us that the confidence level of the result is better than 99%. I generally consider a grade of B- or better as denoting successful students so this can motivate a slightly different comparison of treatment and control groups. In an average UC Davis Physics class of this kind, 52% of the students would get B- or better. Using this measure for the control group (lecture section II) one finds that a student getting 86 or better on the final exam is doing B- or better work. With this number we can characterize the gain by the treatment group in another way by calculating that (after controlling for the same variables as above) we find that a student in the treatment group was almost twice (1.90 ± 0.55 times) as likely as a student in the control group to receive a B- or better, that is, to be academically successful on this exam.

We also gave the FCI as a posttest at the end of the quarter and found that the two classes both had respectable[3] normalized gains in the FCI, 0.38 ± 0.06 for treatment and 0.34 ± 0.06 for control. Though the treatment group had higher conceptual learning gain the result is not statistically significant at the 95% confidence level. This is, perhaps, not surprising since both sections I and II would be classified[3] as interactive engagement classes.

**Conclusions**

One of the important ways to decide whether a specific teacher or a specific teaching method might improve the learning of our student is by doing a controlled study. In this paper I have outlined the general characteristics of two classes that, offered together, constitute a controlled study of a particular organization of curricular materials. The results from this study strongly suggest that students should spend time on qualitative understanding of problems well before undertaking the computationally-complicated quantitative problems. Since the treatment and control classes both devoted a considerable amount of in-class time to student-student discussions of qualitative problems, this study cannot decide whether interactive engagement[3] (via student-student discussion) is needed in order to improve problem solving abilities. However, there is considerable research in physics education that shows that these kinds of in-class discussions promote student understanding of qualitative physics issues so it seems likely that the student-student discussions are necessary. Finally, because this kind of separation of into qualitative and quantitative is usually possible in physics classes, the author will probably always teach this way in the future.

---

[1] D. Hestenes, M. Wells, and G. Swackhamer, "Force Concept Inventory", The Physics Teacher, 30, 141-158 (1992).

[2] E. Kim and S. J. Pak, "Students do not overcome conceptual difficulties after solving 1000 traditional problems", American Journal of Physics, 70, 759-765, (2002).

[3] R. R. Hake, "Interactive-engagement versus traditional methods: A six-thousand-student survey of mechanics test data for introductory physics courses", American Journal of Physics, 66, 64-74, (1998).

[4] Eric Mazur, *Peer Instruction: A User's Manual* (Prentice–Hall, Upper Saddle River, NJ, 1997)

[5] From Chap.1 of: E. F. Taylor and J. A. Wheeler, *Spacetime Physics* (W. H. Freeman, San Francisco, CA, 1966).

[6] D. J. Webb, "Improving student's problem-solving ability as well as conceptual understanding without sacrificing the physics content of a class", submitted to American Journal of Physics.